\documentclass{moriond}
\usepackage{overpic, subfig}

\def\Journal#1#2#3#4{{#1} {\bf #2}, #3 (#4)}

\def\PLB{{\em Phys. Lett.}  B}
\def\PRL{\em Phys. Rev. Lett.}
\def\PRD{{\em Phys. Rev.} D}

\def\CPC{{\em Chin. Phys.} C}

\def\EPJC{{\em Eur. Phys. J.} C}

\def\JHEP{\em JHEP}

\def\NC{\em Nature Commun.}

\def\be{\begin{equation}}
\def\ee{\end{equation}}
\def\bea{\begin{eqnarray}}
\def\eea{\end{eqnarray}}

\newcommand{\Photo}{}

\begin{document}
\vspace*{4cm}
\title{Recent results of (semi)leptonic decays \\of charm hadrons at BESIII}

\author{Xiang Pan\footnote{email:panxiang@suda.edu.cn} \\(For BESIII Collaboration)}

\address{Soochow University, Suzhou 215006, People’s Republic of China}

\maketitle
\abstracts{
In this talk, we report the recent measurements of (semi)leptonic decays of charm mesons from the BESIII experiment, including $D^+\to\mu^+\nu_\mu$, $D^+\to\tau^+\nu_\tau$, $D^+\to\eta^\prime\ell^+\nu_\ell$, and $D^{0(+)}\to\bar K\ell^+\nu_\ell$ (where $\ell=e, \mu$). These measurements provide the most precise or first determinations to date of the CKM matrix elements $|V_{cs(d)}|$, the decay constant $f_{D^+}$, and the hadronic form factors $f_+^{D^+\to \eta^\prime}(0)$ and $f_+^{D\to \bar K}(0)$. Lepton flavor universality of $e-\mu$ and $\tau-\mu$ are also tested with these decays. Additionally, we present the first observation of the rare beta decay $\Lambda_c^+\to ne^+\nu_e$ with machine learning of Graph Neural Network.
}

\section{Introduction}

The Cabibbo-Kobayashi-Maskawa (CKM) matrix, which is a fundamental element of the Standard Model (SM), can only be determined through experimental measurements. The CKM matrix describes the quark mixing between the flavor eigenstates and weak interaction eigenstates during weak interactions. The SM requires that the CKM matrix elements satisfy unitarity, ensuring that quarks exist in only three generations. Therefore, precise measurements of CKM matrix elements are crucial for testing the unitarity of the CKM matrix, thereby testing the SM and probing potential new physics.
Currently, the uncertainties of the CKM matrix elements are dominated by the $|V_{cs}|$ and  $|V_{cd}|$, with uncertainties of 0.6\% and 1.8\%, respectively~\cite{ParticleDataGroup:2024cfk}. The (semi)leptonic (SL) decays of charm mesons offer good opportunity for determining the CKM matrix elements $|V_{cs}|$ and  $|V_{cd}|$.

The couplings between the three families of leptons and the gauge bosons are expected to equal in the SM. This property is known as lepton flavor universality (LFU). In recent years, however, hints of tensions between experimental measurements and the SM predictions were reported in the SL $B$ decays~\cite{HFLAV}.
Precision tests of LFU in different SL decays of heavy mesons provide deeper insight into these anomalies.

BESIII experiment has collected $e^+e^-$ collision data samples of 20.3 $\rm fb^{-1}$, 7.33 $\rm fb^{-1}$,  and 4.5 $\rm fb^{-1}$ at 3.773 GeV, 4.128-4.226 GeV,  and 4.6-4.7 GeV, respectively. These datasets offer the world's largest samples of charm hadrons near their production threshold, including $D^{0(+)}$, $D_s^+$, and $\Lambda_c^+$. These datasets provide a unique opportunity to determine the CKM matrix elements$|V_{cs}|$ and $|V_{cd}|$, as well as to test LFU in the charm sector.

\section{Study of leptonic decay $D^+\to\ell^+\nu_\ell$}

In the SM of particle physics, 
 the partial width of $D^+\to \ell^+\nu_\ell$  can be written as~\cite{Silverman:1988gc}
\begin{equation}
\Gamma_{D^+\to\ell^+\nu_\ell}=\frac{G_F^2}{8\pi}|V_{cd}|^2f^2_{D^+}m_\ell^2 m_{D^+} \left (1-\frac{m_\ell^2}{m_{D^+}^2} \right )^2,
\label{eq1}
\end{equation}
where
$G_F$ is the Fermi coupling constant,
$f_{D^+}$ is the $D^+$ decay constant, $m_\ell$ is the lepton mass, and $m_{D^+}$ is the $D^+$ mass.
Using the lifetime of $D^+$ meson ($\tau_{D^+}$) and the measured branching fractions (BFs) of $D^+\to\ell^+\nu_\ell$, one can determine the product $f_{D^+}|V_{cd}|$. By taking $f_{D^+}=(212.1\pm0.7)$ MeV from recent lattice quantum chromodynamics (LQCD) calculations~\cite{FlavourLatticeAveragingGroupFLAG:2021npn} as input, the value of $|V_{cd}|$ can be obtained. Alternatively, $|V_{cd}|=0.22486\pm0.00067$ can be taken from the global fits of the other CKM matrix elements~\cite{ParticleDataGroup:2024cfk}, and $f_{D^+}$ can be determined instead.

Previously, the measurements of $D^+\to\mu^+\nu_\mu$ have been performed by MARKIII~\cite{Adler:1987ty}, BES~\cite{BES:1998iue}, BESII~\cite{BES:2004ufx}, CLEO~\cite{CLEO:2004pwu,CLEO:2005jsh,CLEO:2008ffk}, and BESIII~\cite{BESIII:2013iro}.
Additionally, the observation of $D^+\to\tau^+\nu_\tau$ via $\tau^+\to\pi^+\bar\nu_\tau$ has been reported by BESIII~\cite{BESIII:2019vhn}. 
We herein present the improved  measurements  of $D^+\to\mu^+\nu_\mu$~\cite{BESIII:2024kvt} and $D^+\to\tau^+\nu_\tau$~\cite{BESIII:2024vlt} with larger data samples. The measured BFs, $f_{D^+}$, and $|V_{cd}|$ are summarized in Table~\ref{tab:lep}. 
The most accurate results of $f_{D^+}$ and $|V_{cd}|$ are given by Ref.~\cite{BESIII:2024kvt}, corresponding to precision of 1.2\%. Combining with the BFs of $D^+\to\mu^+\nu_\mu$~\cite{BESIII:2024kvt} and $D^+\to\tau^+\nu_\tau$~\cite{BESIII:2024vlt}, the ratio $\mathcal R_{\tau/\mu}$ is determined to be 2.49$\pm$0.31, which is consistent with the LFU prediction of $\mathcal R^{\rm SM}_{\tau/\mu}=\frac{m_\tau^2(1-m_\tau^2/m_{D^+}^2)^2}{m_\mu^2(1-m_\mu^2/m_{D^+}^2)^2}=2.69$ within uncertainty.

\begin{table}[htbp]\centering
\caption{Results of the BFs, $f_{D^+}$, and $|V_{cd}|$. The first, second, and third uncertainties are statistical, systematic, and input, respectively.\label{tab:lep}}
  \begin{tabular}{c|c|c|c|c}
  \hline
Channel&$\mathcal L$ ($\rm fb^{-1}$)& BF($\times10^{-4}$)&   $f_{D^+}$ (MeV)  &$|V_{cd}|$ \\
  \hline
$\mu^+\nu_\mu$~\cite{CLEO:2008ffk}&0.818&$3.82\pm0.32\pm0.09$&$207\pm9\pm2\pm1$&$0.220\pm0.009\pm0.003\pm0.001$\\
$\mu^+\nu_\mu$~\cite{BESIII:2013iro}&2.93&$3.71\pm0.19\pm0.06$&$204\pm5\pm2\pm1$&$0.216\pm0.006\pm0.002\pm0.001$\\
$\mu^+\nu_\mu$~\cite{BESIII:2024kvt}&20.3&$3.98\pm0.08\pm0.04$&$212\pm2\pm1\pm1$&$0.224\pm0.002\pm0.001\pm0.001$\\
\hline
$\tau^+\nu_\tau$~\cite{BESIII:2019vhn}&2.93&$12.0\pm2.4\pm1.2$&$225\pm23\pm11\pm1$&$0.238\pm0.024\pm0.012\pm0.001$\\
$\tau^+\nu_\tau$~\cite{BESIII:2024vlt}&7.9&$9.9\pm1.1\pm0.5$&$204\pm11\pm5\pm1$&$0.216\pm0.012\pm0.006\pm0.001$\\

     \hline
  \end{tabular}
\end{table}

\section{Study of $D^{0(+)}\to P\ell^+\nu_\ell$ decay dynamics}

In SM, the differential decay rate of  $D^{0(+)}\to P\ell^+\nu_\ell$ is written as~\cite{Faustov:2019mqr}

\begin{equation}
       \frac{d\Gamma}{dq^2}=
  \frac{G_{F}^{2}|V_{cd}|^{2}}{24\pi^{3}}
  \frac{(q^{2}-m^{2}_{\ell})^2|\vec{p}_{P}|}{q^{4}m^{2}_{D}}
  \left [ \left( 1+\frac{m^{2}_{\ell}}{2q^{2}} \right)
     m^{2}_{D} |\vec{p}_{P}|^2 |f_{+}(q^{2})|^{2}    
  + \frac{3m^{2}_{\ell}}{8q^{2}}
  \left( m^{2}_{D}-m^{2}_{P} \right)^{2}
  |f_{0}(q^{2})|^{2}\right ],
\end{equation}
where $\vec{p}_{P}$ is the three-momentum of the pseudoscalar meson $P$ in the rest frame of the $D^{0(+)}$ meson, and $f_{+,0}(q^2)$ represents the hadronic form factors of the hadronic weak current that depend on the square of the four-momentum transfer $q=p_{D^{0(+)}}-p_P$. These form factors(s) describe strong interaction effects that can be calculated in LQCD. 

The hadronic FF, $f_{+,0}(q^2)$, is parameterized with the two-parameter series expansion~\cite{Becher:2005bg}, in which the  product of $f_{+}(0)|V_{cd}|$ and the shape parameter $r_1$ are to be determined.
Similar formulas are applied for $f_0(q^2)$ but with a one-parameter series expansion, due to the small contribution from this term, and with the pole mass taken from the nearest scalar charm meson.

\subsection{$D^0\to K^-\ell^+\nu_\ell$ and $D^+\to \bar K^0\ell^+\nu_\ell$}
In the past decades, the BFs of $D^0 \to K^-\ell^+\nu_\ell$ and
$D^+ \to \bar K^0\ell^+\nu_\ell$ were measured by
BESII~\cite{BES:2004rav,BES:2004obp,BES:2006kzp},
BaBar~\cite{BaBar:2007zgf}, Belle~\cite{Belle:2006idb},
CLEO-c~\cite{CLEO:2005rxg,CLEO:2005cuk,CLEO:2007ntr,CLEO:2009svp}, and
BESIII~\cite{BESIII:2021mfl,BESIII:2015tql,BESIII:2018ccy,BESIII:2017ylw,BESIII:2016hko,BESIII:2015jmz,BESIII:2016gbw}.
Studies of the decay dynamics of $D \to \bar K\ell^+\nu_{\ell}$ were reported by BaBar~\cite{BaBar:2007zgf}, CLEO-c~\cite{CLEO:2009svp}, and BESIII~\cite{BESIII:2015tql,BESIII:2018ccy,BESIII:2017ylw,BESIII:2015jmz}.
The previous BESIII analysis used 2.93 fb$^{-1}$ of
$e^+e^-$ collision data taken at $\sqrt
s=3.773$~GeV.  This work~\cite{BESIII:2024slx} reports the improved measurements of the
BFs and decay dynamics of $D^0\to K^- \ell^+\nu_\ell$
and $D^+\to \bar K^0 \ell^+\nu_\ell$ by using 7.93~fb$^{-1}$ of
$e^+e^-$ collision data collected at $\sqrt
s=3.773$ GeV~\cite{Luminosity}.

The obtained BFs are 
$\mathcal B(D^0\to K^-e^+\nu_e)=(3.521\pm0.009_{\rm stat.}\pm0.016_{\rm syst.}) \%$, 
$\mathcal B(D^0\to K^-\mu^+\nu_\mu)=(3.419\pm0.011_{\rm stat.}\pm0.016_{\rm syst.}) \%$, 
$\mathcal B(D^+\to \bar K^0e^+\nu_e)=(8.864\pm0.039_{\rm stat.}\pm0.082_{\rm syst.}) \%$, and 
$\mathcal B(D^+\to \bar K^0\mu^+\nu_\mu)=(8.665\pm0.046_{\rm stat.}\pm0.084_{\rm syst.}) \%$, respectively.
Combining the BFs of semielectronic and semimuonic decays, the
ratios of the two BFs are determined to be $\frac{\mathcal
B_{D^0\to K^-\mu^+\nu_\mu}}{\mathcal B_{D^0\to K^-e^+\nu_e}}=0.971\pm0.004_{\rm stat.}\pm0.006_{\rm
syst.}$ and $\frac{\mathcal B_{D^+\to \bar K^0\mu^+\nu_\mu}}{\mathcal
B_{D^+\to \bar K^0e^+\nu_e}}=0.978\pm0.007_{\rm stat.}\pm0.013_{\rm syst.}$, which are
consistent with the theoretical calculation
$0.975\pm0.001$~\cite{Riggio:2017zwh} based on LFU assumption.

\begin{figure*}[htbp]
\begin{center}
\includegraphics[width=0.7\textwidth]{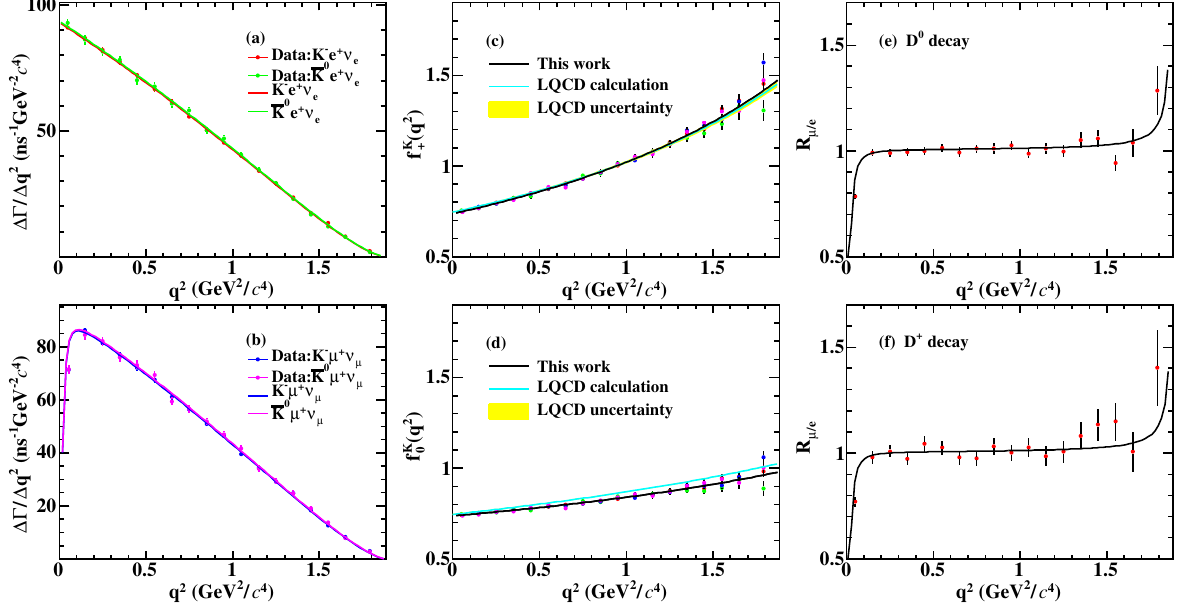}
\caption{(a)(b) Simultaneous fit to partial decay rates of
  $D^0(D^+)\to \bar K\ell^+\nu_\ell$. (c)(d) Projections of
  $f^{K}_+(q^2)$ and $f^{K}_0(q^2)$ as functions of $q^2$ of
  $D^0(D^+)\to \bar K\ell^+\nu_\ell$. (e)(f) The ratio of differential
  decay rates of $D^0\to K^-\mu^+\nu_\mu$ over $D^0\to K^-e^+\nu_e$ and the ratio of differential
  decay rates of $D^+\to \bar K^0\mu^+\nu_\mu$ over $D^+\to \bar K^0e^+\nu_e$ in each $q^2$ bin.}
\label{ff_klnu}
\end{center}
\end{figure*}

From the simultaneous fit to the partial decay rates of $D^0\to K^-e^+\nu_e$,
$D^0\to K^-\mu^+\nu_\mu$, $D^+\to \bar K^0e^+\nu_e$, and $D^+\to \bar K^0\mu^+\nu_\mu$, as shown in Fig.~\ref{ff_klnu}, the product of the hadronic form
factor $f^K_+(0)$ and the modulus of the CKM matrix element $|V_{cs}|$ is
determined to be $f^K_+(0)|V_{cs}|=0.7171\pm0.0011_{\rm
stat.}\pm0.0013_{\rm syst.}$. Taking the value of $|V_{cs}| = 0.97349\pm0.00016$ given by
the PDG~\cite{ParticleDataGroup:2024cfk} as input, we obtain the hadronic form factor
$f^K_+(0)=0.7366\pm0.0011_{\rm stat.}\pm0.0013_{\rm syst.}$.  Conversely,
using the $f^K_+(0)$ calculated in LQCD~\cite{FermilabLattice:2022gku}, we obtain
$|V_{cs}|=0.9623\pm0.0015_{\rm stat.}\pm0.0017_{\rm
syst.}\pm0.0040_{\rm LQCD}$. The measured $f^K_+(0)$ and $|V_{cs}|$ are the most precision results to date in the world.

\subsection{$D^+\to\eta^\prime\ell^+\nu_\ell$}
In 2011 and 2018, CLEO and BESIII reported the BF of $D^+\to\eta^{\prime}e^+\nu_e$~\cite{CLEO:2010pjh,BESIII:2018eom} with large uncertainties, of order $25\%$.  We herein report the first observation of $D^+\to \eta^\prime \mu^+\nu_\mu$, the improved measurement of $D^+\to \eta^\prime e^+\nu_e$, and the first analysis of $D^+\to \eta^\prime \ell^+\nu_\ell$ decay dynamics~\cite{BESIII:2024njj} by analyzing 20.3~fb$^{-1}$~\cite{BESIII:2024lbn} of $e^+e^-$ collision data taken at $\sqrt{s}=$3.773 GeV.

The measured BFs are $\mathcal B(D^+\to\eta^\prime\mu^+\nu_\mu)=(1.92\pm0.28_{\rm stat}\pm0.08_{\rm syst})\times10^{-4}$ and $\mathcal B(D^+\to\eta^\prime e^+\nu_e)=(1.79\pm0.19_{\rm stat}\pm0.07_{\rm syst})\times10^{-4}$. The ratio of $\mathcal B(D^+\to\eta^\prime\mu^+\nu_\mu)$ to $\mathcal B(D^+\to\eta^\prime e^+\nu_e)$ is $\mathcal R_{\mu/e}=1.07\pm0.19_{\rm stat}\pm0.03_{\rm syst}$, which is consistent with the SM predictions of LFU~\cite{Soni:2018adu,Cheng:2017pcq}. 

\begin{figure}
\centering
  \includegraphics[width=0.4\textwidth]{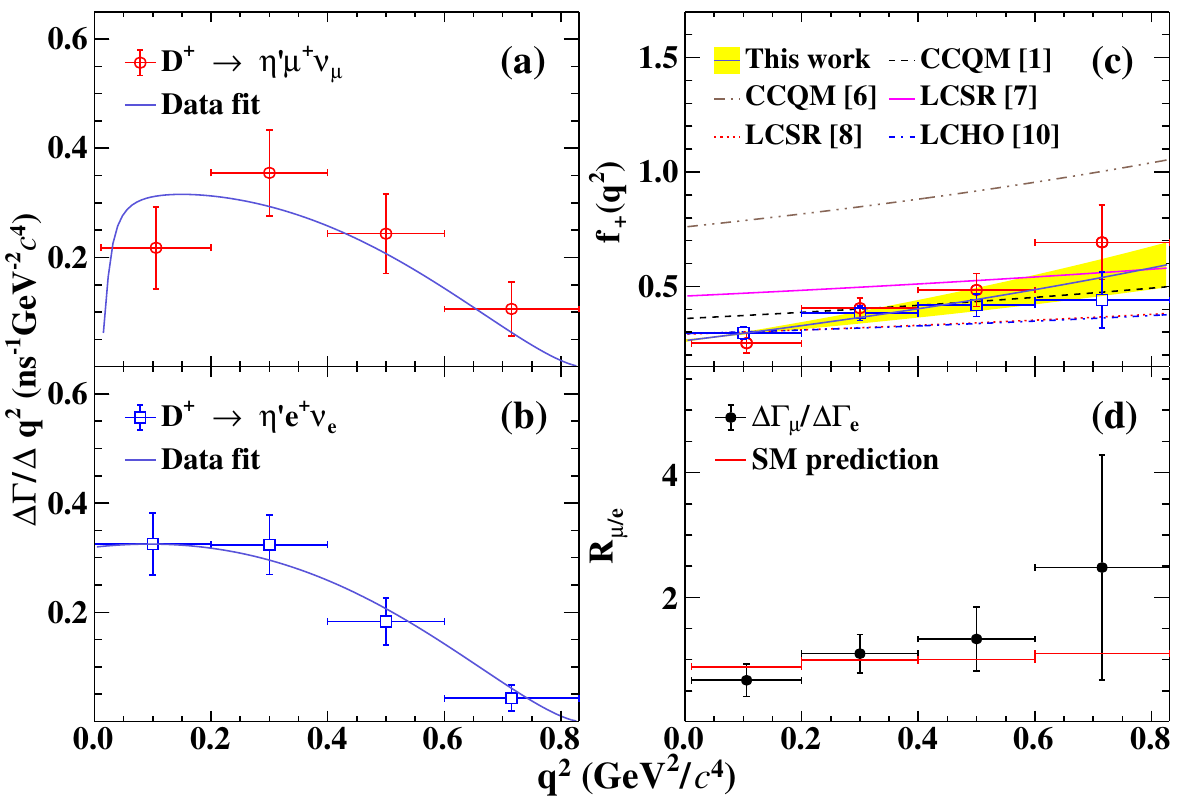}
  \includegraphics[width=0.4\textwidth]{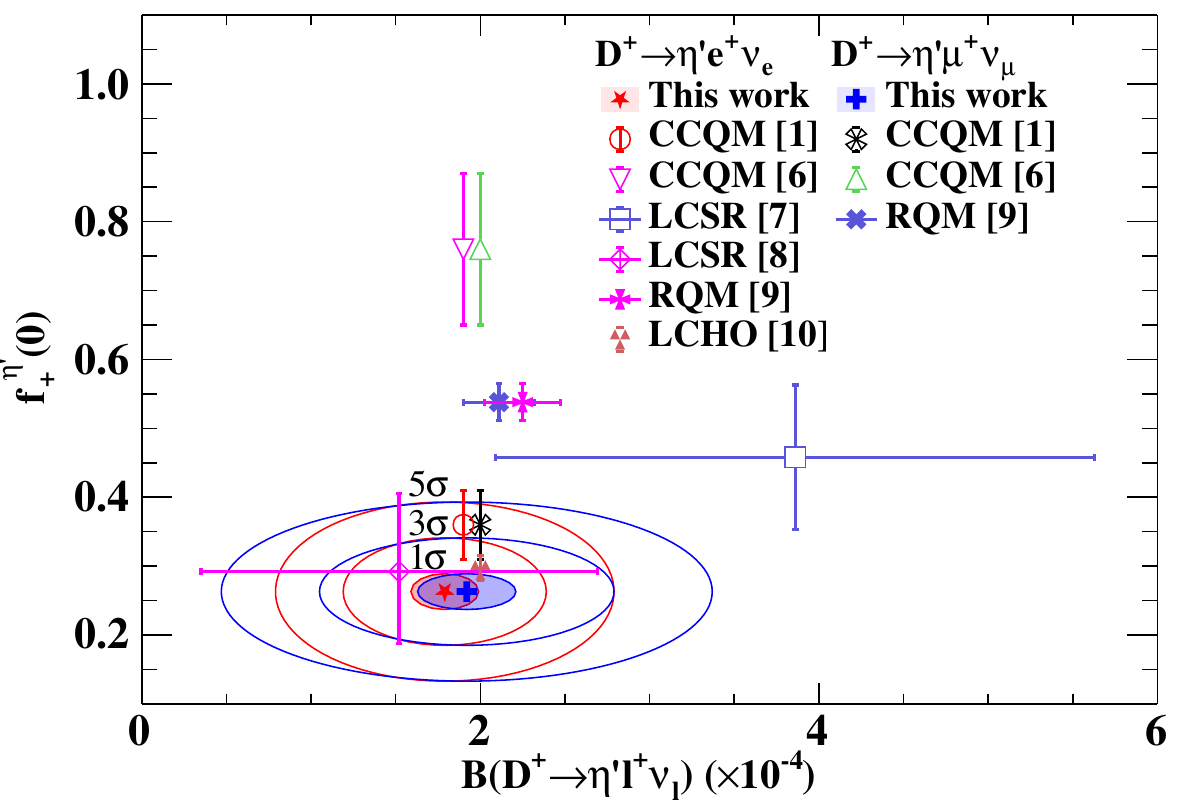}

\caption{\small (a, b) Fits to $\Delta\Gamma^i_{\rm msr}$, (c) projections to $f_+^{\eta^{\prime}}(q^2)$,  (d) the measured $\mathcal R_{\mu/e}$ in each $q^2$ interval, and (e) comparison of the BFs and FFs measured in this work and those given by different theoretical predictions.}
  \label{fig:combine_FF}
\end{figure}

By analyzing the partial decay rates of $D^+\to\eta^\prime\ell^+\nu_\ell$, the FF is determined to be $f^{\eta^{\prime}}_+(0)=0.263\pm0.025_{\rm stat}\pm0.006_{\rm syst}$ for the first time. 
Figure~\ref{fig:combine_FF} shows the fit results and the comparison of the measured and different theoretical predicted BFs and FFs.  
The measured results are important to test different theoretical calculations.

\section{Study of $\Lambda_c^+\to ne^+\nu_e$}
The SL decay of the lightest charmed baryon $\Lambda_c^+$ offers unique insights into the mechanisms of strong and electroweak interactions. 
To date, the Cabibbo-favored SL decays $\Lambda_c^+\to\Lambda e+\nu_e$ and $\Lambda_c^+\to\Lambda \mu+\nu_\mu$ have been well studied at BESIII~\cite{BESIII:2022ysa,BESIII:2023jxv}.
However, the Cabibbo-suppressed decay $\Lambda_c^+\to ne^+\nu_e$ involves two missing particles in the final state, the neutron and the neutrino, causing significant experimental challenges. The neutron is difficult to reconstruct using the Electromagnetic Calorimeter (EMC). Additionally, the dominant background events from $\Lambda_c^+\to\Lambda e^+\nu_e, \Lambda\to n\pi^0$,  exceed the signal yield by a factor of ten, requiring an advanced signal identification method to discriminate the energy deposition patterns of neutrons from those of $\Lambda$ hyperons in the EMC.

\begin{figure}[htp]
    \centering
    \begin{overpic}[height=4.5cm]{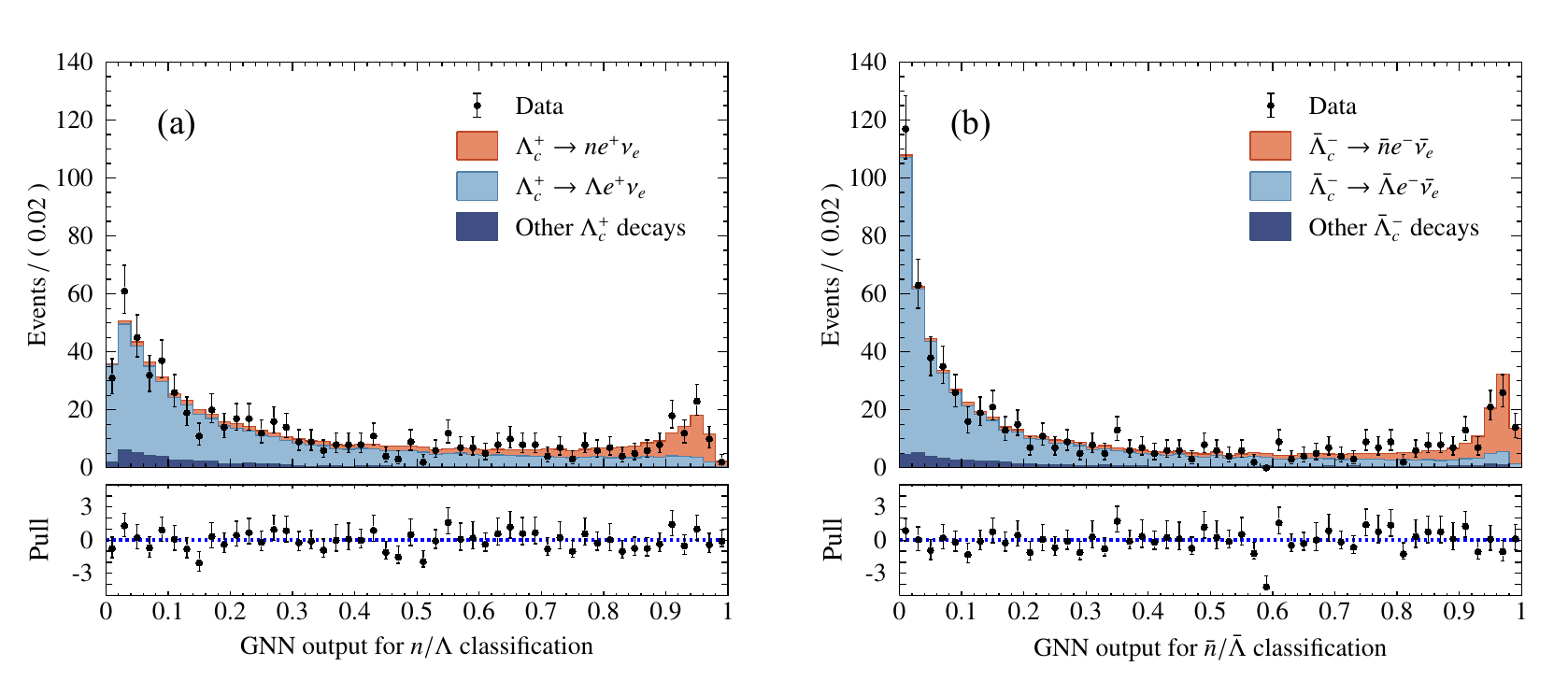}
\end{overpic}
    \begin{overpic}[height=4.5cm]{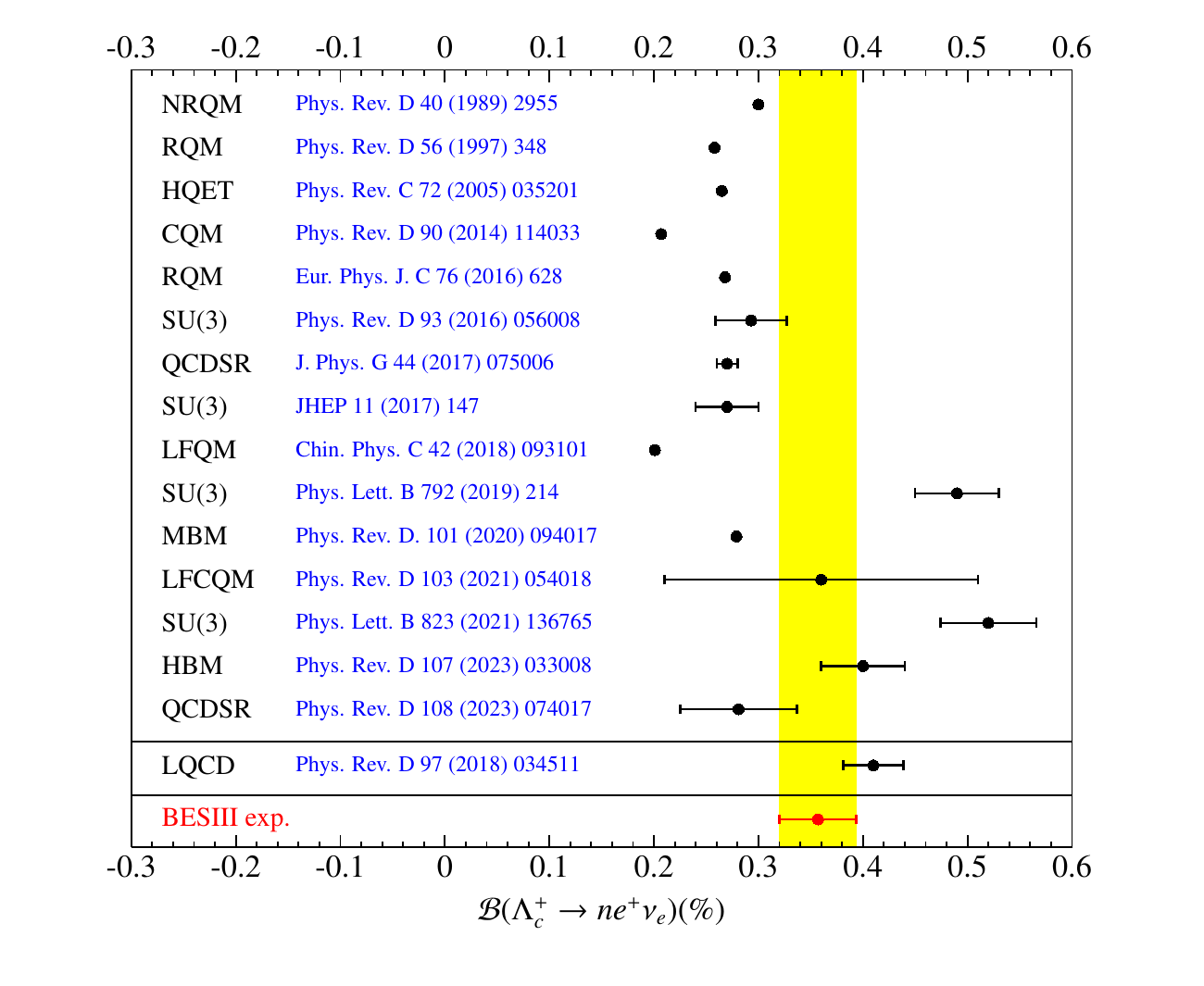}
    \put(75,65){\footnotesize(c)}
\end{overpic}
 \caption{(a) Fit to the GNN output distribution in $\bar{\Lambda}_c^{-}\to\bar{n} e^{-}\bar{\nu}_e$ signal candidates. (b) Fit to the GNN output distribution in $\Lambda_c^{+}\to n e^{+}\nu_e$ signal candidates. (c) Comparison of our BF measurement with the theoretical predictions. }\label{fig:fit}
\end{figure}

In Ref.~\cite{BESIII:2024mgg}, 
the machine learning of Graph Neural Network (GNN) architecture is employed to identificated the neutrons and $\Lambda$. 
 By representing the EMC particle showers as a point cloud and processing it with a GNN architecture, a classification model is trained to separate neutron-like and $\Lambda$-like patterns. 
The $\Lambda_c^+\to ne^+\nu_e$ signal yield is extracted from the GNN output probability distributions, as shown in Figs.~\ref{fig:fit} (a, b), revealing clear enhancements at the high end with a statistical significance exceeding 10 $\sigma$.

Based on 4.5 fb$^{-1}$ data collected at $\sqrt s$ from 4.6 to 4.7 GeV, the absolute BF of $\Lambda_c^+\to ne^+\nu_e$ is measured to be $(0.357\pm0.034_{\rm stat}\pm0.014_{\rm syst})$\%.  Figure~\ref{fig:fit}(c) shows the comparison of our BF measurement with different theoretical predictions. 
A recent LQCD calculation~\cite{Meinel:2017ggx} gives the $q^2$-integrated partial width of $\Lambda_c^+\to ne^+\nu_e$ as
 $\Gamma(\Lambda_c^+\to ne^+\nu_e)=|V_{cd}|^2(0.405\pm0.016\pm0.020)$ ps$^{-1}$. Using the measured BF and current $\Lambda_c^+$ lifetime, the CKM matrix element $|V_{cd}|$ is determined from charmed baryon decays for the first time, yielding $0.208\pm0.011_{\rm stat}\pm0.007_{\rm syst}\pm0.001_{\rm input}$. This work highlights a new approach to further understand fundamental interactions in the charmed baryon sector, and showcases the power of modern machine learning techniques in experimental particle physics.

\section{Summary}
\begin{figure}[htp]
\centering
\includegraphics[height=5.0cm]{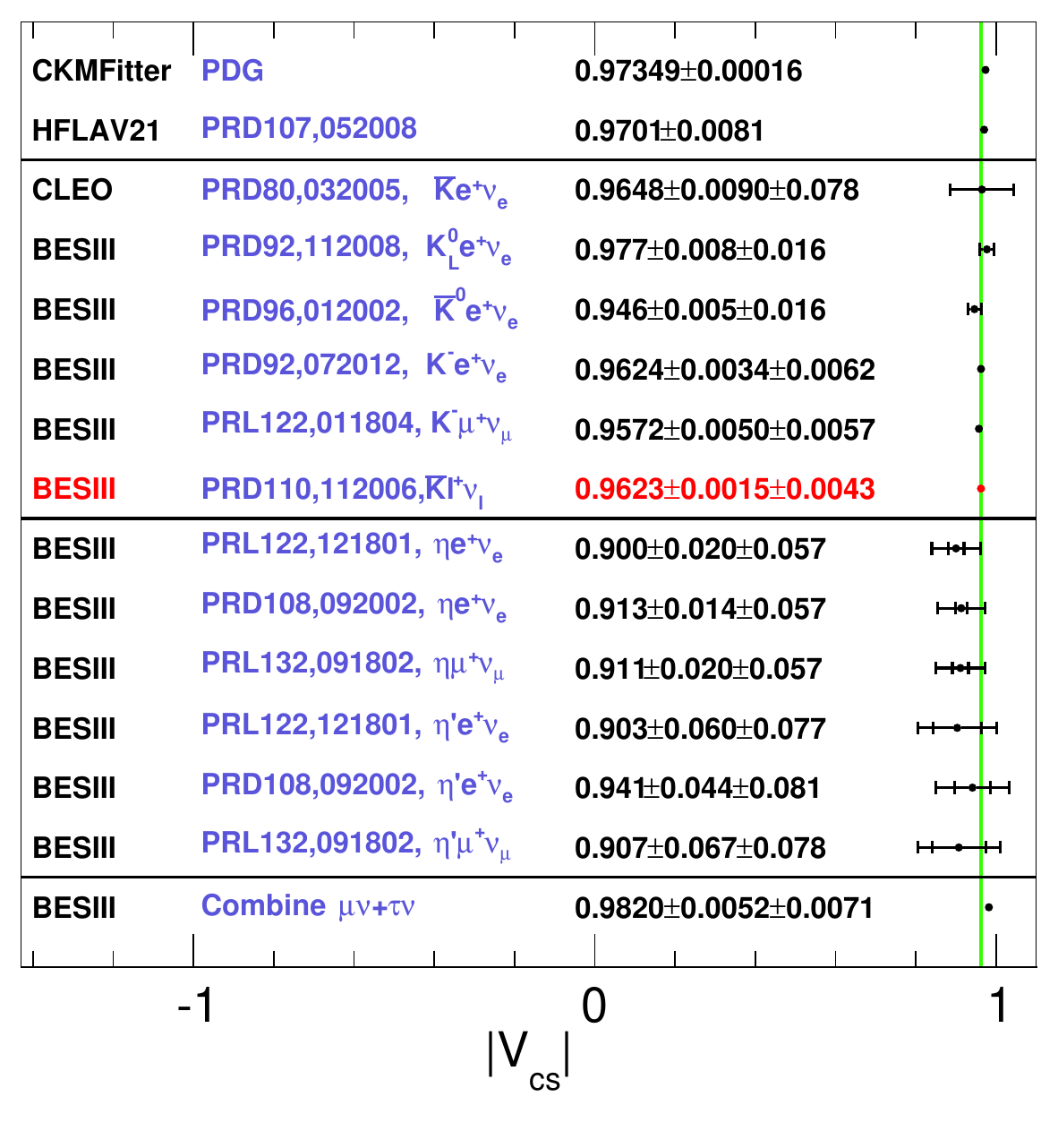}
\includegraphics[height=5.0cm]{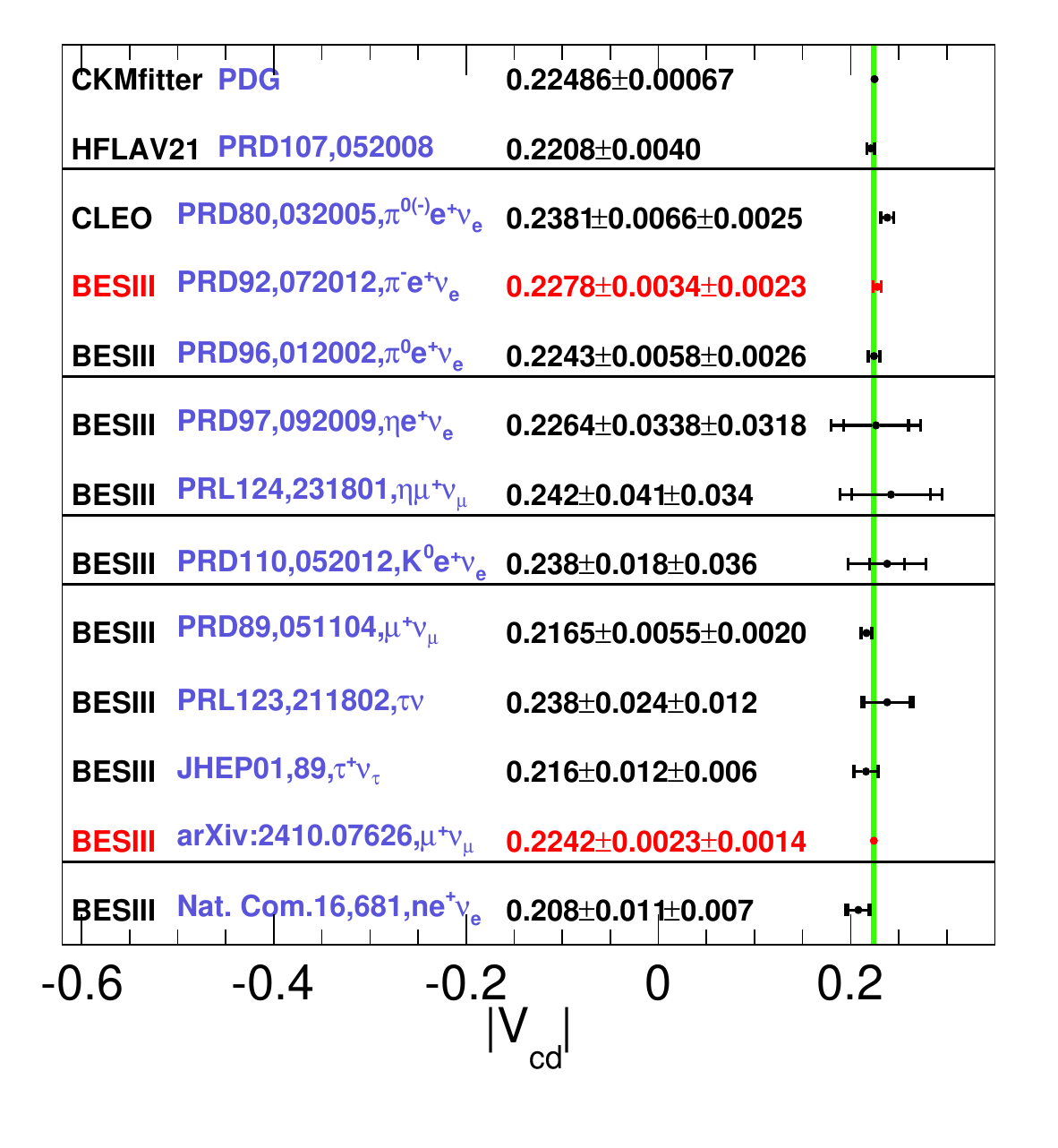}
\includegraphics[height=5.0cm]{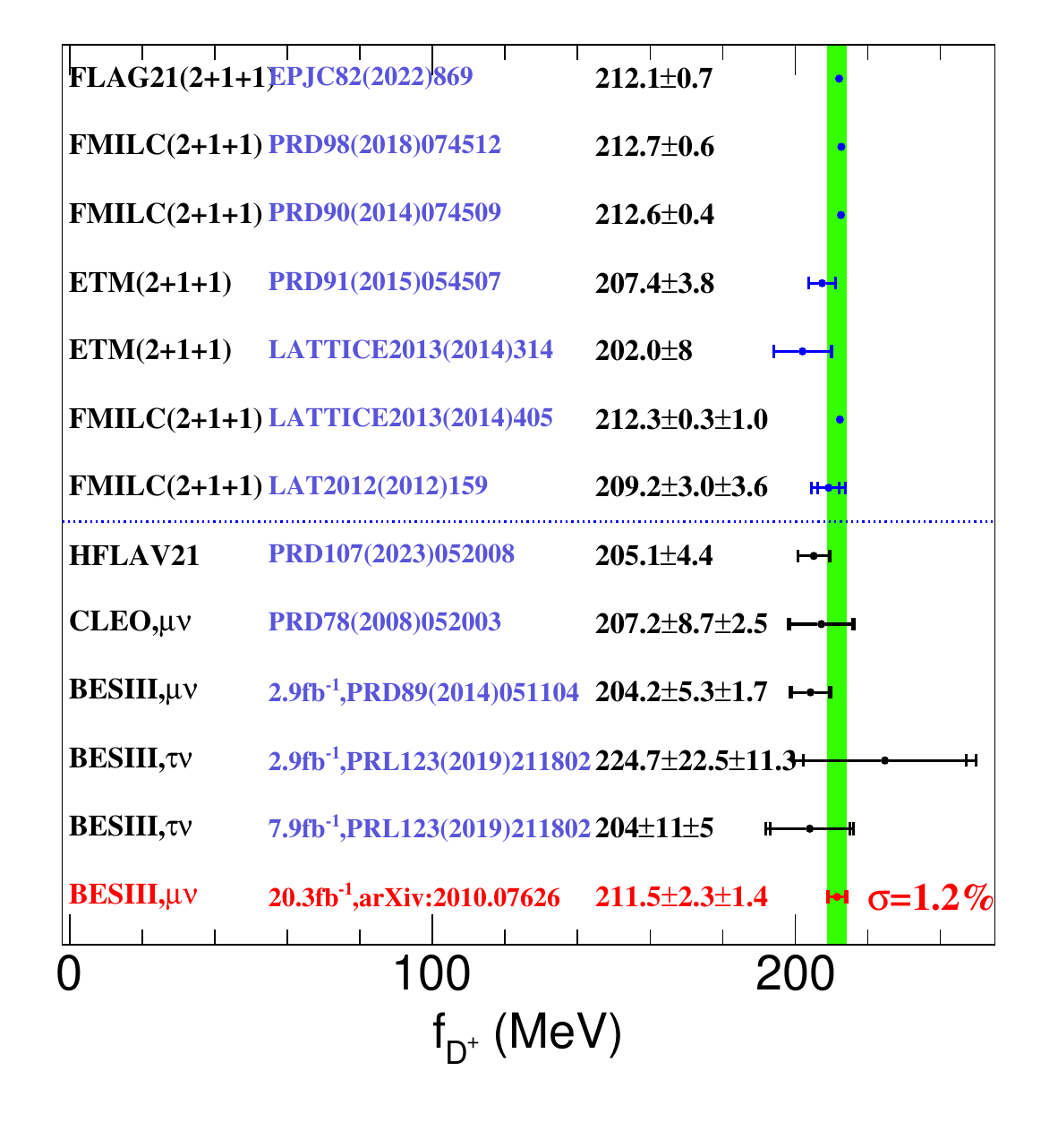}
\includegraphics[height=5.0cm]{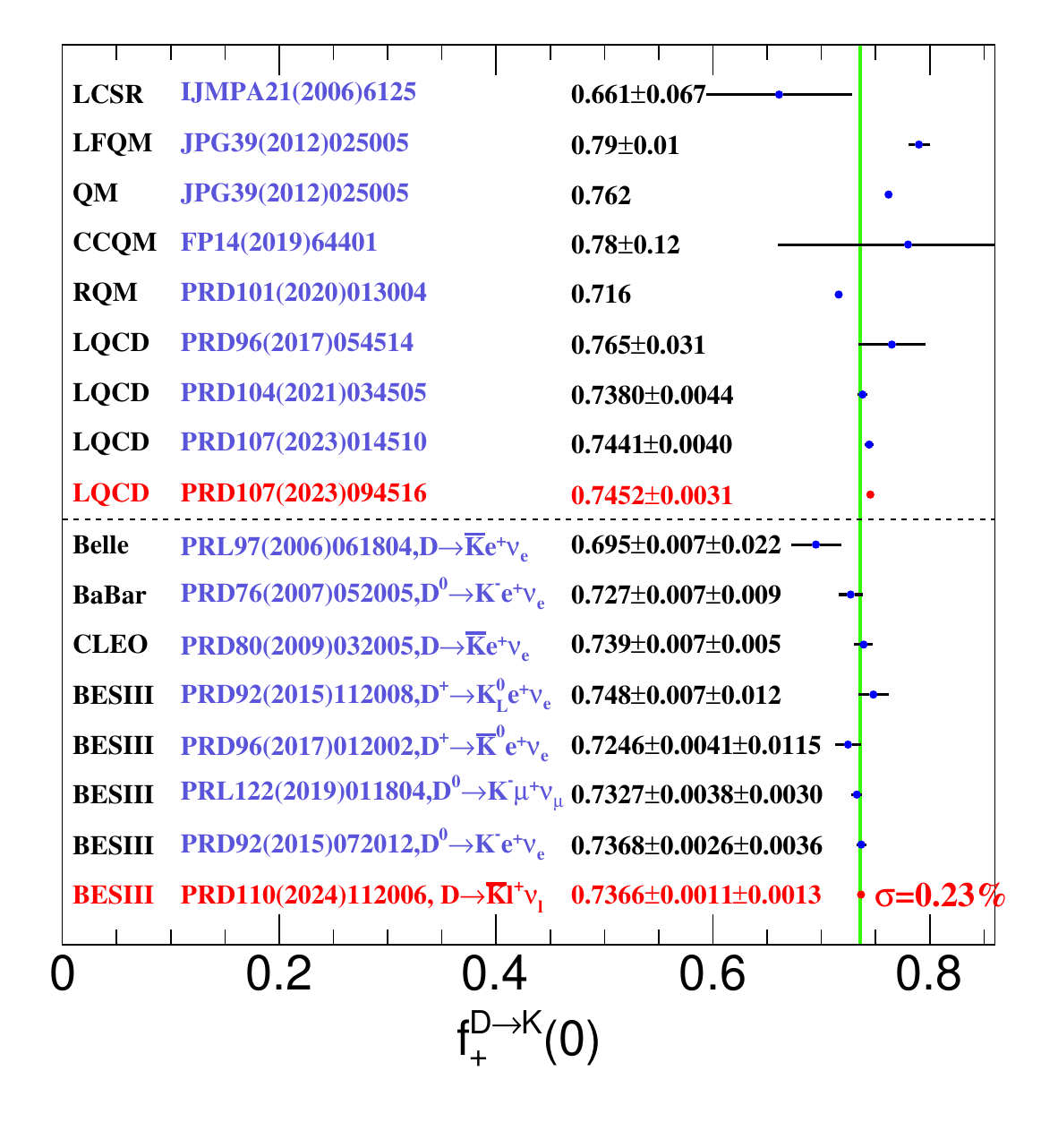}
\includegraphics[height=5.0cm]{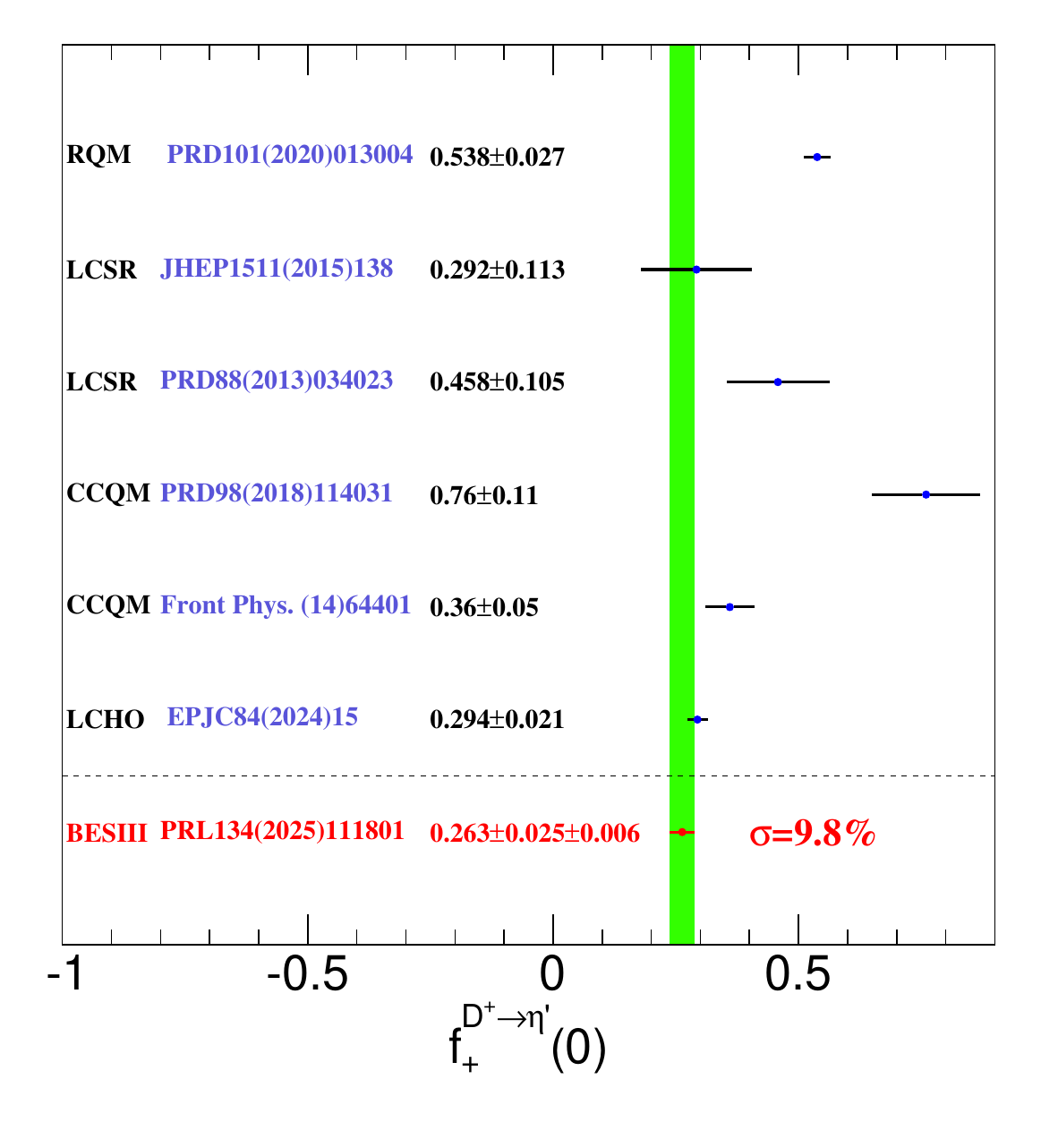}

\caption{Comparisons of $|V_{cs}|$, $|V_{cd}|$, decay constant $f_{D^+}$, and FFs $f_+^{D\to \bar K}(0)$, $f_+^{D\to\eta^\prime}(0)$ determined from  different experiments. For experimental results, the inner error bar is the statistical uncertainty and the outer is the combined statistical and systematic uncertainty. The green band denotes  the most precision experimental result.
\label{compare_vcq}
}
\end{figure}

By analyzing the data samples collected at $D\bar D$ and $\Lambda_c^+\bar\Lambda_c^-$ mass threshold, the SL decays of charmed hadrons have been studied at the BESIII, including the improved measurements of $D^+\to\mu^+\nu_\mu$, $D^+\to\tau^+\nu_\tau$,  $D^{0(+)}\to \bar Ke^+\nu_e$, and  $D^{0(+)}\to \bar K\mu^+\nu_\mu$, and first observation of $D^+\to\eta^\prime\mu^+\nu_\mu$ and $\Lambda_c^+\to ne^+\nu_e$.
Figure~\ref{compare_vcq} compares the $|V_{cs}|$, $|V_{cd}|$, decay constant $f_{D^+}$, and FFs $f_+^{D\to \bar K}(0)$, $f_+^{D\to\eta^\prime}(0)$ determined from different experiments.
 The uncertainties of $|V_{cs}|$, $|V_{cd}|$, $f_{D^+}$ and $f_+^{~D\to \bar K}(0)$ have been reduced to 0.5\%, 1.2\%, 1.2\% and 0.3\%, respectively. No evidence of LFU violation is found within precision of 0.74\% for $\mathcal R(\mu/e)$ in $D\to \bar K\ell^+\nu_\ell$ and 12\% for $\mathcal R(\tau/\mu)$ in $D^+\to\ell^+\nu_\ell$. The observation of $\Lambda_c^+\to ne^+\nu_e$ demonstrates the power of machine learning in experimental particle physics, offering the potential in searching for the rare decays of charm hadrons in future.

BESIII has collected 20.3 fb$^{-1}$ of data at 3.773 GeV and plans to collect an additional  3 fb$^{-1}$ of data at 4.178 GeV and 10 fb$^{-1}$ of data between 4.6-4.7 GeV in future~\cite{BESIII:2020nme}. More precise measurements and searches for rare SL decays will be presented.

\end{document}